\documentclass[12pt,dvips]{article}

\usepackage{amsmath,amssymb,exscale}
\usepackage{array,multicol}
\usepackage{afterpage,float,flafter}
\usepackage{epsfig,rotating,pifont}
\usepackage{cite}

\def\simgt{\stackrel{>}{{}_\sim}}
\makeatletter
\@addtoreset{equation}{section}
\makeatother

\title{
\vspace*{-0.8cm}
\begin{flushright}
\end{flushright}
\vspace{1cm} 
\Large\textbf{``Gaugomaly''  Mediated SUSY Breaking \\
and \\
Conformal Sequestering}
\vspace*{.5cm}
\author{\large \textbf{
Raman Sundrum\footnote{email: ~ sundrum@pha.jhu.edu} 
}\\
\\
\emph{
Department of Physics and Astronomy} \\ 
\emph{Johns Hopkins University} \\ 
\emph{3400 North Charles St}. \\ 
\emph{Baltimore, MD 21218-2686}}}

\date{}
\begin{document}
\maketitle
\thispagestyle{empty}
\vspace*{.5cm}
  
\begin{abstract} 
Anomaly-mediated supersymmetry breaking in the context of 4D conformally 
sequestered models is combined with Poppitz-Trivedi D-type 
gauge-mediation. The implementation of the 
two mediation mechanisms naturally leads to visible 
soft masses at the same scale so that they can cooperatively solve the 
$\mu$ and flavor problems of weak scale supersymmetry, as well as the 
tachyonic slepton problem of pure anomaly-mediation. The tools are 
developed in a modular fashion for more readily fitting into 
the general program of  optimizing supersymmetric dynamics 
in hunting for the most  
attractive weak scale phenomenologies combined with Planck-scale 
plausibility. 
\end{abstract} 
  
\newpage 
\renewcommand{\thepage}{\arabic{page}} 
\setcounter{page}{1}


\section{Introduction}

The scenario of weak scale supersymmetry (SUSY) (reviewed in Ref. 
\cite{review}), while broadly attractive, 
faces several technical challenges: A successful theory should explain 
(i) what mechanism 
robustly protects rare processes such as flavor-changing neutral currents 
(FCNCs) from excessive superpartner-mediated corrections,
(ii) how gauge coupling unification, apparent in the softly 
broken minimal supersymmetric standard model (MSSM), 
is preserved in the full theory, (iii) how visible sector 
SUSY breaking triggers electroweak symmetry breaking, 
and (iv) how all superpartners and Higgs bosons naturally
obtain masses that evade present search constraints.  While economical 
mechanisms are available to address any of these issues individually, 
addressing them collectively is rather difficult. The purpose of this 
paper is to extend the set of mechanisms upon which realistic 
model-building may be based. Indeed a realistic model is presented, although 
it is hoped that further investigation, and perhaps additional mechanisms, 
will yield models with larger acceptable parameter spaces.   

Issue (i), in particular the SUSY flavor problem, is sensitive to the 
mechanism by which SUSY breaking, originating in a hidden sector, 
is mediated  to the visible sector. A promising approach is to have the 
mediating force be flavor-blind, the natural candidates being gauge forces 
or gravity. Gauge-mediated SUSY breaking (GMSB) \cite{gmsb} \cite{gmsb2} (reviewed 
in Ref. \cite{giudice}), 
as well as gaugino-mediated SUSY breaking \cite{ggmsb},
are based on the first 
possibility, while anomaly-mediated SUSY breaking \cite{amsb1} \cite{amsb2} 
is based on the second. 
While GMSB and AMSB are compatible with (ii), they both face significant
 difficulties with (iii) and (iv). Both suffer from the  $\mu$-problem. 
AMSB applied to the MSSM yields tachyonic sleptons \cite{amsb1} 
which destabilize the 
standard electroweak vacuum. 
In this paper, we will 
discuss how GMSB and AMSB can naturally combine symbiotically in 
addressing (i) -- (iv). This general direction is not new. Ref. 
\cite{pomrat} demonstrated within a non-minimal visible sector that 
AMSB effects could trigger competitive GMSB contributions. Ref. \cite{kribkap}
proposed ``gaugino-assisted'' AMSB. Refs. \cite{jones} \cite{kaplan} 
 combined AMSB with SUSY breaking arising 
from gauge D-terms.
Finally, Refs. \cite{chalut} \cite{weiner}, which are 
of particular relevance to the present paper, proposed that
 AMSB could naturally 
combine with the ``D-type'' of GMSB studied by Poppitz and Trivedi \cite{PT}. 

A central issue is that of sequestering, ensuring that the AMSB and GMSB 
contributions to the visible sector are not overshadowed by 
SUSY breaking from Planck-suppressed (superspace) 
contact interactions 
directly with the hidden sector. Originally this was achieved by 
considering the visible and hidden sectors to be separated in a 
minimally inhabited extra dimension \cite{amsb1} \cite{ls1}. 
While this setting is quite 
economical and 
technically natural, it does introduce the 
extra element of non-renormalizable extra-dimensional effective field theory
on top of non-renormalizable 4D (super-)gravity. While string theory 
strongly motivates the existence of extra dimensions, it is not easy to assess
 how readily extra dimensions with the requisite minimal light 
field content arise. Refs. \cite{dine} 
studied string theoretic examples where 
these conditions were violated, but it is still hard to know how generic 
their conclusions are. 

A more recent approach to sequestering is to remain in 4D, but
 dynamically suppress visible-hidden contact interactions through strong 
renormalization effects in running them down from the Planck scale. 
This requires either the visible sector or the hidden sector to be strongly 
coupled over a large hierarchy (and therefore approximately conformal) 
below the Planck scale. Refs. \cite{NS} studied the first possibility in the 
context of gaugino-mediated SUSY breaking. Refs. \cite{confseq1} 
\cite{confseq2}
studied the latter possibility
 as well as the conceptual connections between this 
conformal sequestering and extra-dimensional sequestering (in particular of 
the warped form studied in Ref. \cite{warpseq})
via the 
AdS/CFT correspondence \cite{adscft} \cite{rs} 
\cite{rscft}. The advantage here
is that, when based on known super-conformal field theories (SCFT's), 
there is no 
doubt as to the {\it existence and properties} of the sequestering 
dynamics. 

Conformal sequestering has been criticized recently \cite{fox} 
for its reliance on 
hidden discrete symmetries, or non-genericity in order 
to avoid reliance on these symmetries. It is still unclear, because of 
our relative ignorance of the whole ``landscape'' of strongly coupled SCFT's, 
whether such  aesthetic considerations reflect a serious and generic problem 
with conformal sequestering or our restriction thusfar to simple controllable 
examples.

The present paper will employ the conformal sequestering 
dynamics of Ref. \cite{confseq2}, used there 
in a pure AMSB context, but extended here with a messenger sector that 
leads to competitive GMSB contributions to the visible 
sector\footnote{The particular hidden sector model in Ref. \cite{confseq1}
 depends on a certain strong interaction coefficient being negative, whereas
its sign can in fact be determined to be positive by chiral lagrangian 
symmetry arguments. The model can be remedied, and this will be detailed 
elsewhere. We proceed here with Ref. \cite{confseq2} which avoids this 
problem.}. The results particularly resemble those of the extra-dimensionally 
sequestered models of Ref. \cite{chalut}, although the mechanisms are quite 
distinct (even after mapping qualitative features via the AdS/CFT 
correspondence). It should also be mentioned that realistic models employing 
conformal sequestering, combining AMSB and $D$-terms 
of gauged $B-L$ and hypercharge 
were developed in Ref. \cite{murayama}.

To appreciate the level of theoretical control, note that
the effective interactions presented below do employ non-renormalizable 
(but 4D) Planck-suppressed interactions, which are natural to expect 
along with gravity. But, with the obvious exception of (super-)gravity,
 all of the interactions can be made renormalizable at the expense of 
introducing some extra massive fields. We will not do this here. The 
important physical scales will lie far below $M_{Pl}$, so that only 
classical supergravity is important. That is, the level of control is 
similar to doing QED in the earth's gravitational field. 


The organization of this paper is as follows.
The salient features of Ref. \cite{confseq2} will be reviewed in Section 2. 
A compatible messenger sector will be added in Section 3, explaining 
how GMSB is naturally competitive with AMSB. Section 4 will present a 
realistic model (except that we have not bothered with neutrino masses) 
based on having the visible sector be the next-to-minimal SUSY standard 
model (NMSSM). Section 5 discusses extensions aimed at broadening the  
parameter space in different directions, chiefly to give 
maximal freedom to the energy scales for the {\it origins} of visible 
flavor structure, while still solving the supersymmetric flavor problem.

\section{Hidden Sector Dynamics}

We will adopt the hidden sector construction of Ref. \cite{confseq2}, 
which we therefore briefly review in this section.

\subsection{Dynamical SUSY breaking}

The hidden sector comprises two SQCD sub-sectors with Planck-suppressed 
couplings between them. 
One of the sub-sectors, $SQCD_2$, has two colors and four 
``flavors'' (eight doublets), denoted for convenience 
by $T_{Ia}, I = 1,2,3,4, a = 1,2$. The theory lies at the 
self-dual point of Seiberg's conformal window, and flows to a 
strongly-coupled  IR fixed point. We assume that at the Planck scale this 
subsector is already near this fixed point. 
The other sub-sector, $SQCD_3$, has three colors and two flavors, 
$P_a, \overline{P}_a, ~ a = 1,2$, and is weakly coupled at the Planck scale. 
The hidden superpotential is given by 
\begin{equation}
W_{hid} = \frac{\lambda'}{M_{Pl}}~\sum_{I \neq J} 
(T_{aI} T^{a}_I) (T_{bJ} T^{b}_J) 
+ \frac{\lambda}{M_{Pl}} \sum_I (T_{aI} T^{a}_I) (\overline{P}_a 
P^a), 
\end{equation}
where $\lambda, \lambda' \sim {\cal O}(1)$, the parantheses 
denote  gauge-invariant bilinears, and the $a$ index is 
raised using the $\epsilon_{ab}$-tensor. 

There are regions of the {\it classical} hidden moduli space where it is 
one(-complex)-dimensional, parametrized by an $SQCD_2$ quark bilinear, 
$(T_{1a} T^{a}_1)$. However, non-perturbative effects lift and 
stabilize this modulus. First, because of the non-trivial fixed point 
dynamics  
of the $SQCD_2$ subsector, the canonically normalized modulus is 
given by $X \equiv (T_{1a} T^{1a})^{2/3}/M_{Pl}^{1/3}$, that is the 
effective Kahler potential at the fixed point is $\sim X^{\dagger} X$ 
for $X \ll M_{Pl}$. Secondly, non-perturbative effects from the 
$SQCD_3$ sub-sector, acting through $\lambda$ dynamically 
generate a linear superpotential for $X$. The net result is an 
effective Polonyi model of SUSY breaking \cite{polonyi} 
by $X$, where $\langle X \rangle
\ll M_{Pl}$ (unlike the original Polonyi model) 
is stabilized by a combination of SUGRA corrections and 
the leading Kahler corrections for $X$ (determined by the flow to the 
non-trivial fixed point). It should be noted that the vacuum so obtained is 
only metastable, but can easily be stable on cosmological time scales.

The hidden SUSY breaking is given by 
\begin{equation}
F_X \equiv \Lambda_{int}^2,
\end{equation}
where the intermediate scale $\Lambda_{int} \ll M_{Pl}$ naturally due 
to its non-perturbative origins. Once the cosmological constant is cancelled
by inclusion of a constant superpotential, one finds
\begin{equation}
F_{SUGRA} \sim m_{3/2} \sim  \frac{F_X}{M_{Pl}} = 
\frac{\Lambda_{int}^2}{M_{Pl}}.
\end{equation}

\subsection{Conformal sequestering, hidden symmetries and AMSB}

Visible sector SUSY breaking can now arise by couplings to 
 $F_{SUGRA}$ and $F_X$. The SUGRA couplings of the visible sector 
ensure that SUSY breaking inherited from $F_{SUGRA}$ is proportional to the 
breaking of visible conformal invariance.
For a classically scale-invariant visible sector with the standard 
loop-level scale anomaly, the visible SUSY breaking pattern is given by 
AMSB \cite{amsb1} \cite{amsb2}. 
To compare later with gauge-mediated soft terms let us note that the 
 visible AMSB soft masses are all of the rough  form, 
\begin{equation}
m_{soft} \sim (\alpha_{vis}/4 \pi) F_{SUGRA},
\end{equation}
a visible loop factor being necessary in order to be sensitive to the quantum 
scale anomaly.

Visible SUSY breaking from $F_X$ goes through effective superspace 
operators which mix 
the hidden sector with the visible sector. Such effects can dominate over 
AMSB and ruin  the solution it offers to the 
SUSY flavor problem. Some of these effective visible-hidden operators 
(in particular superpotential terms) can 
naturally be avoided by appealing to 
non-renormalization theorems. The most dangerous unprotected 
operators  are Kahler terms of the form
\begin{equation}
\Delta K \sim \frac{Q^{\dagger}_i Q_j T^{\dagger}_{Ia} T_{Jb}}{M_{Pl}^2}, 
\end{equation}
where $i, j$ label visible flavors and $I, J, a, b$ label hidden flavors.
In the IR, $T^{\dagger}_{Ia} T_{Jb}$ 
becomes proportional to $X^{\dagger} X$ and results in the direct 
communication of hidden SUSY breaking to the visible sector. Since the 
couplings have no reason to preserve visible flavor symmetry (already broken 
in the standard model), visible flavor will be violated by SUSY breaking 
originating from $F_X$. 

 The central subtlety in estimating 
how $T^{\dagger}_{Ia} T_{Jb}$ reduces to $X^{\dagger} X$ in the IR is
 taking account of the 
strong renormalization effects of the $SQCD_2$ dynamics as the operator 
is run down from the Planck scale down to $\langle X \rangle \ll M_{Pl}$ 
(which spontaneously 
breaks the approximate conformal invariance of the $SQCD_2$ 
sub-sector).  Let us decompose the hidden bilinear 
into a hidden-flavor singlet part and a non-singlet part (containing no 
singlet), 
\begin{equation}
T^{\dagger}_{Ia} T_{Jb} = (T^{\dagger}_{Ia} T_{Jb})_{non-singlet} + 
(T^{\dagger}_{Ia} T_{Jb})_{singlet}.
\end{equation}
Now the non-singlet piece is a non-anomalous hidden flavor current 
(super-multiplet) and therefore is not renormalized by the $SQCD_2$ 
dynamics. Since $X^{\dagger} X$ is not pure singlet, couplings of
the form 
\begin{equation}
\Delta K \sim \frac{Q^{\dagger}_i Q_j ~(T^{\dagger}_{Ia} 
T_{Jb})_{non-singlet}}{M_{Pl}^2},
\end{equation}
translate in the IR to 
\begin{equation}
\Delta K_{eff} \sim \frac{Q^{\dagger}_i Q_j X^{\dagger} X}{M_{Pl}^2},
\end{equation}
implying flavor-violating SUSY breaking 
for visible scalars,
\begin{equation}
m_{scalar}^2 \sim \frac{|F_X|^2}{M_{Pl}^2} = \frac{\Lambda_{int}^4}{M_{Pl}^2}.
\end{equation}
 
Such large flavor-violating SUSY breaking would make AMSB irrelevant and 
introduce the SUSY flavor problem. Fortunately, the visible coupling 
to the hidden non-singlet can be forbidden by insisting on some amount 
of hidden flavor symmetry. It is easy to check that $W_{hid}$ is 
compatible with having exact dynamical symmetries comprising the 
discrete symmetries, 
\begin{eqnarray}
\label{discrete}
(i) ~ T_{1a} &\leftrightarrow&  T_{2a}, ~ T_{3a} \leftrightarrow  T_{4a} \nonumber \\ 
(ii) ~ T_{1a} &\leftrightarrow&  T_{3a}, ~ T_{2a} \leftrightarrow  T_{4a} \nonumber \\
(iii) ~ T_{Ia} &\rightarrow& - T_{Ia}, ~ T_{Jb} \rightarrow T_{Jb}, ~ 
J \neq I, 
\end{eqnarray}
as well as an $SU(2)$ symmetry acting on all $a$-indices\footnote{
This $SU(2)$ symmetry can either be a global symmetry, or to protect 
it from gravitational corrections, it can be very weakly gauged. 
Alternatively, one only needs a non-abelian discrete subgroup of the 
$SU(2)$.}. This 
symmetry is powerful enough to accidentally forbid the visible 
couplings to the non-singlet hidden bilinear. 

On the other hand, there is no exact dynamical symmetry that can forbid the 
 visible sector couplings to the hidden flavor singlet bilinear,
\begin{equation}
\Delta K \sim \frac{Q^{\dagger}_i Q_j |T|^2}{M_{Pl}^2},
\end{equation} 
simply because $|T|^2$ 
is of the form of the $T$ kinetic term itself. The whole point of 
conformal sequestering is that $|T|^2$ 
corresponds to an anomalous current, and has an order one 
anomalous dimension suppressing the mixed coupling in the running between 
$M_{Pl}$ and $\langle X \rangle$, quite plausibly enough to render AMSB 
so dominant as to solve the SUSY flavor problem.

\section{Messenger Sector}

We now add a messenger sector to facilitate GMSB. 

\subsection{Fields, superpotential and hidden symmetries}

The messenger sector consists of new chiral fields charged under 
standard model gauge symmetry, $\psi, \chi, \overline{\psi}, \overline{\chi}$
and singlets $A, B, C, D$. In order to maintain the unification of couplings 
of the SUSY standard model, under the embedding of the standard model gauge 
group into $SU(5)$, we take $\psi, \chi$ to be $5$'s and 
$\overline{\psi}, \overline{\chi}$ to be $\overline{5}$'s. 
The messenger sector is distinguished from the visible sector in that, while 
the visible sector does not transform under the hidden discrete symmetries, 
Eq. (\ref{discrete}), the messenger fields transform non-trivially under (i), 
\begin{equation}
(i) ~ \chi \leftrightarrow \psi, ~ ~ 
\overline{\chi} \leftrightarrow \overline{\psi}, 
~ ~ A \leftrightarrow B, ~ ~ C \rightarrow C, ~ ~ D \rightarrow D,
\end{equation}
but are inert under (ii), (iii).

The messenger superpotential (compatible with the hidden discrete symmetries) 
is given by
\begin{eqnarray}
W_{mess.} &=& \lambda_{+} (\overline{\chi} \chi + 
\overline{\psi} \psi)(A + B) ~ + ~ 
\lambda_{-} (\overline{\chi} \chi - 
\overline{\psi} \psi)(A - B) \nonumber \\ 
&+&  \lambda_C C(A^2 + B^2 - v_{mess}^2) ~ + ~
\lambda_D D A B,
\end{eqnarray}
with the dimensionless $\lambda$ couplings again being order one.
The mass parameter $v_{mess} \ll M_{Pl}$ is treated here as an 
 input scale although it is  simple to realize the small 
ratio $v_{mess}/M_{Pl}$ as arising from naturally small non-perturbative 
effects\footnote{For example, a pure SUSY Yang-Mills sector coupling to 
$C$ according to $\int d^2 \theta ~C {\cal W}_{\alpha}^2/M_{Pl}$ will 
generate a non-perturbative 
linear superpotential (and negligible higher order terms) 
for $C$ upon gaugino-condensation.}. 

$F$-flatness in the purely singlet part of the superpotential results in 
the VEVs
\begin{equation}
A = v_{mess}, ~ ~ B = C = D = 0,
\end{equation}
with all masses $\sim v_{mess}$. That is, the hidden 
discrete symmetry (i) is broken in the IR by $v_{mess}$. This results in 
messenger mass terms  of order $v_{mess}$,
\begin{equation}
W_{mess.-mass} = (\lambda_{+} +  \lambda_{-}) v_{mess} ~\overline{\chi} \chi
+ (\lambda_{+} -  \lambda_{-}) v_{mess} ~\overline{\psi} \psi,
\end{equation}
which do not respect the hidden discrete symmetry (i). We will see why this 
is important in the next section. 

It seems possible that the messenger and hidden sectors can be more 
economically combined, in particular since the messenger singlets are, 
with some contrivance, spontaneously 
breaking hidden discrete symmetries which are 
already broken by the hidden sector. However, in the interests of modularity 
this will not be pursued here.

\subsection{Poppitz-Trivedi Gauge Mediation}

Our analysis now 
will be similar to some of the steps taken in Ref. \cite{chalut}.
Just as with the visible sector, there are Kahler couplings of the 
messengers to the hidden flavor singlet current, such as
\begin{eqnarray}
\Delta K &\sim& \frac{(\chi^{\dagger} \chi + \psi^{\dagger} \psi)
(|T_{1a}|^2 + |T_{2a}|^2 + |T_{3a}|^2 + |T_{4a}|^2)}{M_{Pl}^2}, \nonumber \\
&~& \frac{|C|^2
(|T_{1a}|^2 + |T_{2a}|^2 + |T_{3a}|^2 + |T_{4a}|^2)}{M_{Pl}^2},
\end{eqnarray}
but just as in the visible case, these are conformally sequestered and 
insignificant. However, 
the non-trivial transformation of the messengers under (i) also allows 
one to write  Kahler couplings to  hidden flavor 
non-singlet currents, such as 
\begin{eqnarray}
\label{messK}
\Delta K &\sim&  \frac{(\chi^{\dagger} \chi - \psi^{\dagger} \psi)
(|T_{1a}|^2 - |T_{2a}|^2 + |T_{3a}|^2 - |T_{4a}|^2)}{M_{Pl}^2},
\end{eqnarray}
which is not renormalized by the strong 
$SQCD_2$ dynamics, and matches to an IR coupling of the form
\begin{equation}
\Delta K_{eff} \sim \frac{(\chi^{\dagger} \chi - \psi^{\dagger} \psi) 
X^{\dagger} X}{M_{Pl}^2}.
\end{equation}
This yields the following soft mass-squareds for messenger scalars upon 
picking out the $F_X^{\dagger} F_X$ term\footnote{It is straightforward to 
check that all terms involving the lowest component VEV $\langle X \rangle 
\ll M_{Pl}$ lead to subdominant effects. Of course there are messenger 
soft terms arising from $F_{SUGRA}$ as well, but the resulting contributions 
to visible soft masses is  
additive to gauge-mediation, and because of the UV insensitivity of the 
former, is subsumed in the
 AMSB effects at the weak scale. These, having been 
reviewed in the previous section, are neglected in this section.}:
\begin{equation}
\label{seed}
m_{\overline{\chi}soft}^2 = - 
m_{\overline{\psi}soft}^2 \sim m_{\chi soft}^2 
= - m_{\psi soft}^2 \sim \frac{\Lambda_{int}^4}{
M_{Pl}^2}.
\end{equation}

It is this soft SUSY breaking carried by the messengers which is  
mediated  to the visible sector scalars by their shared
 gauge interactions at two loops, in the manner determined by Poppitz and 
Trivedi \cite{PT}:
\begin{eqnarray} 
\label{PT}
m_{vis.~soft}^2 &\sim& (\alpha_{gauge}/4 \pi)^2 
~ \ln|\frac{\lambda_+  + 
\lambda_-}{\lambda_+ - \lambda_-}| ~ m_{\chi soft}^2 \nonumber \\
&\sim& (\alpha_{gauge}/4 \pi)^2 ~
\ln|\frac{\lambda_+ + 
\lambda_-}{\lambda_+ - \lambda_-}| ~ \frac{\Lambda_{int}^4}{M_{Pl}^2}.
\end{eqnarray}

Note several significant features of the gauge-mediated 
contributions to the visible soft terms. 
{\it The gauge mediated visible soft masses are 
automatically parametrically 
of the same size as AMSB contributions}, namely 
visible-loop-factor$\times \Lambda_{int}^2/M_{Pl}$. 
The gauge-mediated  contributions are to visible scalar mass-squareds only, 
not to 
fermion masses, A-terms or B-terms. 
 The sign and the precise magnitude of the 
contributions is 
a free parameter from the effective field theory point of view, being 
determined by the independent coefficient of the 
hidden non-singlet term in Eq. (\ref{messK}) (which explains why only 
estimates have been given for the soft terms). Thus gauge-mediated 
contributions can naturally solve the tachyonic-slepton problem of minimal 
AMSB by choice of this coefficient \cite{chalut}. 
The 2-loop relation Eq. (\ref{PT}) is UV finite, potential log divergences 
cancelling because of the suppressed messenger supertrace, suppressed 
because of conformal sequestering. The log in Eq. (\ref{PT}) is therefore 
automatically order one, so that no large logs upset the balance of 
gauge-mediated and AMSB visible soft terms. The whole purpose of 
the messenger sector singlets was to spontaneously 
break the hidden discrete symmetries in the IR, allowing $\lambda_- \neq 0$, 
so that the log does not vanish.

The messenger supersymmetric 
mass scale, $v_{mess}$ does not appear explicitly in Eq. (\ref{PT}) (having 
cancelled out of the ratio of messenger masses in the log), so it does not 
have to be tuned with respect to the weak scale, in principle it 
can lie within a very large range. We will put an upper bound on it below 
so that it does not spoil sequestering in the visible sector. 
Another consideration is that gauge mediation only solves 
the SUSY flavor problem if the origins of visible flavor structure itself
lie at scales {\it above} $v_{mess}$. In this paper we are not committing to 
a theory of how such flavor structure arises in the visible sector, but only 
mechanisms to avoid the SUSY flavor problem. Therefore the most 
conservative position is to take as low a $v_{mess}$ as is consistent 
with other model building constraints, so as to allow for the most ``room'' 
in the UV for flavor structure to arise through unspecified dynamics.

\subsection{Other dangerous channels}

Even though the discrete hidden symmetry 
is only broken in the IR, the IR-sensitivity of
 Poppitz-Trivedi mediation (logarithmic in the messenger 
mass) means the log factor represents an order one breaking of the symmetry. 
This is an important point which should be compared with another a priori 
dangerous SUSY-breaking mediation channel. In addition to Planck-suppressed 
visible-hidden and messenger-hidden Kahler terms, one can also have 
visible-messenger couplings such as
\begin{equation}
\Delta K =  \frac{Q^{\dagger}_i Q_j (\chi^{\dagger} \chi + \psi^{\dagger} \psi)}{M_{Pl}^2}.
\end{equation}
Note that the messenger fields must form an invariant combination 
under (i) since the 
visible fields are. Starting with 
 such an interaction and integrating out a  
messenger loop, threatens to mediate the SUSY breaking in Eq. (\ref{seed}) to 
the visible sector. The diagrams are UV quadratically divergent and yield 
visible {\it flavor-violating} soft mass-squareds 
$\sim \Lambda_{int}^4 \Lambda_{UV}^2/(16 \pi^2 M_{Pl}^4)$. For 
$M_{Pl} < \Lambda_{UV} < 4 \pi M_{Pl}$ this would disasterously dominate 
over AMSB and GMSB contributions. However, the sum of all such divergent 
diagrams cancel because of the discrete symmetry (i), since visible 
fields are even, while the unsequestered messenger mass-squareds are 
purely odd  (originating
 from the spontaneous breaking of the discrete symmetry by the 
hidden sector VEVs). 

Of course, since the messengers also have (i)-violating 
supersymmetric masses, there will be less divergent contributions to visible 
mass-squareds which do not cancel. But they will be suppressed by the 
messenger masses precisely because highly UV divergent diagrams are 
IR insensitive, unlike the Poppitz-Trivedi contributions. That is, 
they will yield visible soft mass-squareds of order 
$\sim \Lambda_{int}^4 v_{mess}^2/(16 \pi^2 M_{Pl}^4)$.
For the resulting (generally flavor-violating) SUSY breaking to be as 
suppressed as the conformally sequestered contributions in Ref. 
\cite{confseq2},
 we must have 
\begin{equation}
\label{bound}
\frac{v_{mess}}{M_{Pl}} < 8 \times 10^{-3}. 
\end{equation}

There are also a priori dangerous contributions from Kahler couplings of all 
three sectors, such as 
\begin{equation}
\Delta K \sim \frac{Q^{\dagger}_i Q_j (|A|^2 - |B|^2) (|T_{1a}|^2 - |T_{2a}|^2 
+ |T_{3a}|^2 - |T_{4a}|^2)}{M_{Pl}^4}.
\end{equation}
(The lower dimension operator with single powers of $A$ and $B$ could be 
forbidden by symmetries $A, B, \chi, \overline{\psi} \rightarrow - A, B, 
\chi, \overline{\psi}$ respectively, with other fields inert.)
Again, the hidden flavor non-singlet current is not renormalized so that in 
the IR this will match onto an operator
\begin{equation}
\Delta K \sim \frac{Q^{\dagger}_i Q_j v_{mess}^2 X^{\dagger} X}{M_{Pl}^4}.
\end{equation}
Again, this is sufficiently suppressed if Eq. (\ref{bound}) is satisified.

\section{Realistic Example}

Ref. \cite{chalut} engineered effective field theories in
 flat 5D compactification with the same interplay of AMSB and 
Poppitz-Trivedi gauge mediation which we have realized here through 
(non-perturbative) 4D gauge dynamics. It is important to stress that the 
mechanisms are not in any sense AdS/CFT dual to each other, indeed the 
proposal of Ref. \cite{chalut} does not extend to (truncated) 5D 
AdS spacetimes. 
The proposal of the present paper is an independent mechanism demonstrating 
how AMSB and gauge mediation can naturally combine. 

However, the phenomenological results of 
Ref. \cite{chalut} can be carried over to the present paper. 
Ref. \cite{chalut} chose the visible sector to consist of just the  NMSSM, 
with the hope that the singlet would develop a VEV to act as the $\mu$-term.
If AMSB by itself dominated visible SUSY breaking, while all soft 
masses would be of the same parametric size 
$\sim$ loop-factor $\times \Lambda_{int}^2/M_{Pl}$, the signs do not 
work out. Sleptons obtain tachyonic mass-squareds, 
while the singlet does not, and hence does not develop a weak scale VEV. 
Running from the Planck scale cannot help because of the characteristic 
UV insensitivity of AMSB. Fortunately, the gauge-mediated contributions 
can correct the sign of the slepton mass-squareds and are  
sensitive to running from the UV. The singlet can then be driven to 
condense at the weak scale and solve the $\mu$-problem.

Ref. \cite{chalut} found fully realistic examples in parameter space by 
treating the gauge-mediated soft Higgs mass-squareds as independent 
from the gauge-mediated soft slepton mass-squareds. This is only possible 
at or above the GUT scale where the Higgs and slepton gauge interactions 
can be distinct, which requires that the messenger scale $v_{mess} 
\simgt M_{GUT}$. This realistic example carries over to the present paper. 
However, note that the requirement $v_{mess} 
\simgt M_{GUT}$ is only just consistent with the requirement of 
Eq. (\ref{bound}).

\section{Extensions}

Let us consider why we might wish to pursue extensions of our results so far.
While a visible NMSSM sector allows one
to find realistic and natural 
SUSY and electroweak symmetry breaking by combining 
AMSB and gauge-mediation, as pointed out in Ref. \cite{chalut} such examples 
are difficult to find in the visible parameter space.
  Also, it is rather restrictive that 
$v_{mess}$ is constrained from above and below to be about the GUT scale in 
order for the Higgs and slepton gauge-mediated 
soft mass-squareds to be independent. This restricts visible sector 
flavor structure to originate in the narrow range of energies between the 
GUT and Planck scales. 
Finally, 
given the mechanism demonstrated here in conformal sequestering 
that gauge-mediation and AMSB can fruitfully compete, it is worthwhile 
understanding more general forms of the gauge-mediated soft terms before 
committing to a particular visible sector model.

It is a straightforward to extend our visible sector from the NMSSM to 
include a $U(1)'$ gauge group, where the charge is given by a linear 
combination of hypercharge and $B-L$ which commutes with $SU(5)$, which
 does not interfere with standard model gauge unification.  To 
cancel anomalies one includes the right-handed neutrino supermultiplet, and 
to eliminate this gauge group by observable energies one includes fields that 
can Higgs the $U(1)'$. We can then add extra messengers which are 
standard model singlets but are vectorially charged under $U(1)'$, with 
their own Planck-suppressed couplings to the hidden sector. This will result 
in an independent Poppitz-Trivedi gauge mediation contribution to the
 visible sector, but now proportional to powers of $U(1)'$ charges, rather 
than standard model charges. (See Refs. 
\cite{affleck} \cite{moha} \cite{bogdan} for earlier proposals of 
$B-L$ mediation.) In this way the gauge mediated contributions to 
the Higgs mass-squareds are independent 
 of the gauge mediated contributions to 
the squarks and sleptons even when $v_{mess} \ll M_{GUT}$. 
The only requirement on the 
 $U(1)'$ breaking scale is that it be below $v_{mess}$. Still, this can be 
at such a high scale that none of the extra $U(1)'$-related structure need 
survive to the weak scale, 
where we can have just the NMSSM. The lower $v_{mess}$, the 
larger the range of energies at which visible sector flavor structure can 
emerge.

 The basic mechanisms decribed here may be 
compatible with a variety of visible sectors. Which of these most naturally 
accomodates the data with minimal fine-tuning at the weak scale 
remains an important and 
ongoing investigation. In future work the possibility of naturally combining 
AMSB with ``F-type'' GMSB will be explored.


\section*{Acknowledgements}

I am very grateful to Kaustubh Agashe, 
David E. Kaplan, Markus Luty and Tsutomu Yanagida
for helpful discussions. This work was supported by NSF Grant PHY-0099468.


\end{document}